\documentclass[twocolumn,preprintnumbers,amsmath,amssymb]{revtex4}
\catcode`\@=11

\def\eqlt{\mathrel{\mathpalette\@vereq<}}  % < over =
\def\eqgt{\mathrel{\mathpalette\@vereq>}}  % > over =
\def\@vereq#1#2{\lower2.5pt\vbox{\baselineskip0pt \lineskip-.5pt
 \ialign{$\m@th#1\hfil##\hfil$\crcr#2\crcr{=}\crcr}}}

\newcommand{\simle}{\ \raise.3ex\hbox{$<$}\kern-0.8em\lower.7ex\hbox{$\sim$}\ }
\newcommand{\simge}{\ \raise.3ex\hbox{$>$}\kern-0.8em\lower.7ex\hbox{$\sim$}\ }

\newcommand{\mib}[1]{\mbox{\boldmath $#1$}}

\begin{document}
%\draft
\title {Nonmagnetic Insulating States near the Mott Transitions on Lattices
with Geometrical Frustration and Implications for $\kappa$-(ET)$_2$Cu$_2$(CN)$_3$}  
\author {  Hidekazu Morita, Shinji Watanabe and Masatoshi  Imada }  
\address { Institute for Solid State Physics, University of Tokyo, Kashiwanoha,
Kashiwa, Chiba, 277-8581, Japan }  
%\date { February 7, 2002 }  
%\maketitle

\begin{abstract} 
We study phase diagrams of the Hubbard model on anisotropic triangular lattices,
which also represents a model for $\kappa$-type BEDT-TTF compounds.  
In contrast with mean-field predictions, path-integral renormalization 
group calculations
show a universal presence of nonmagnetic insulator sandwitched
by antiferromagnetic insulator and paramagnetic metals.  
The nonmagnetic phase does
not show a simple translational symmetry breakings such as 
flux phases, implying a genuine Mott insulator.
We discuss possible relevance on the nonmagnetic insulating phase 
found in $\kappa$-(ET)$_2$Cu$_2$(CN)$_3$.
\end{abstract}
%\pacs{71.30.+h, 71.20.Rv, 71.10.Fd, 75.10.Jm, 71.10.Hf} 
\maketitle
{
%\kword { strongly correlated electron, Hubbard model, 
%metal-insulator transition, magnetic transition, Mott transition, Mott insulator, 
%quantum phase transition, spin liquid, geometrical frustration, triangular lattice,
%BEDT-TTF compound }  
%\begin{document} 
%\sloppy \maketitle  
When the kinetic and Coulomb repulsion energies are severely competing,
the ground states of many-body electron systems can be highly nontrivial.
The Mott transition provides a typical example of such nontrivial behavior.
If the electronic orbitals 
constituting the band near the Fermi level form geometrically frustrated 
lattice structure, the ground state can be even more nontrivial,
because the spin entropy is not easily released by simple symmetry breaking
such as the antiferromagnetic (AF) order, and the possiblity of some exotic quantum
 liquid state opens.
Theoretically, regular triangular lattice has been proposed as such a typical 
structure~\cite{Anderson}, while the nature of the ground 
state of the Hubbard model on 
this lattice still remains a challenge because of the lack of theoretical tools. 
Reliability of mean field approaches becomes questionable under strong quantum 
fluctuations while the numerical studies have been limited to small clusters
because of the lack of efficient algorithm.

In this letter, we obtain the ground-state phase diagram of the Hubbard model on anisotropic
triangular lattice called $t$-$t'$-TH model 
by using recently developed path-integral renormalization group (PIRG) method~\cite{KashimaImadaPIRG}
with systematic inclusion of quantum fluctuations. 
In the $t$-$t'$-TH model, when one of the transfers ($t'$) on a unit triangle becomes different from
the other transfers $t$ as in Fig.~\ref{Fig1}, the model continuously connects the right triangular 
lattice ($t'/t=1$) to the regular square one ($t'/t=0$). 
%In the latter end point at $t'/t=0$ a Mott insulating state with an AF order is believed to be stabilized at any $U \neq 0$.
We study this general model at half filling.

%
%              %%%%%%%    %%%%%        %%%%%%    %%%%%   %     %
%              %      %  %             %     %  %     %   %   %
%              %      %  %             %     %  %     %    % %
%              %%%%%%%    %%%%%        %%%%%%   %     %     %
%              %               %       %     %  %     %    % %
%              %               %       %     %  %     %   %   %
%              %         %%%%%%        %%%%%%    %%%%%   %     %
%
%              By Jean Orloff
%              Comments & suggestions by e-mail: ORLOFF@surya11.cern.ch
%              No modification of this file allowed if not e-sent to me.
%
% A simple way to measure the size of encapsulated postscript figures
%   from inside TeX, and to use it for automatically formatting texts
%   with inserted figures. Works both under Plain TeX-based macros
%   (Phyzzx, Harvmac, Psizzl, ...) and LaTeX environment.
% Provides exactly the same result on any PostScript printer provided
%   the single instruction \psfor... is changed at the end of this
%   file to fit the needs of the particular dvi->ps translator used.
%
% History:
%   1.2.4: fix error handling & add \psonlyboxes
%   1.2.3: adds \putsp@ce for OzTeX fix
%   1.2.2: makes \drawingBox \global for use in Phyzzx
%   1.2.1: accepts %%BoundingBox: (atend)
%   1.2: tries to add \psfordvitps for the TeXPS package.
%   1.1: adds \psforoztex, error handling...
%2345678 1 345678 2 345678 3 345678 4 345678 5 345678 6 345678 7 3456789
%
\catcode`\@=11
% Every macro likes a little privacy...
%
%Trying to tame the variety of \special commands for Postscript: the
%  universal internal command \PSspeci@l##1##2 takes ##1 to be the
%  filename and ##2 to be the integer scale factor*1000 (as for usual
%   TeX \scale commands)
%
\def\psfortextures{%     For TeXtures on the Macintosh
%-----------------
\def\PSspeci@l##1##2{%
\special{illustration ##1\space scaled ##2}%
}}
\def\psfordvitops{%      For the DVItoPS converter on IBM mainframes
%----------------
\def\PSspeci@l##1##2{%
\special{dvitops: import ##1\space \the\drawingwd \the\drawinght}%
}}
\def\psfordvips{%      For DVIPS converter on VAX, UNIX and PC's
%--------------
\def\PSspeci@l##1##2{%
%    \special{/@scaleunit 1000 def}% never read dox without trying!
\d@my=0.1bp \d@mx=\drawingwd \divide\d@mx by\d@my%
\includegraphics{##1\space}%
}}
\def\psforoztex{%        For the OzTeX shareware on the Macintosh
%--------------
\def\PSspeci@l##1##2{%
\special{##1 \space
      ##2 1000 div dup scale
      \putsp@ce{\number-\psllx} \putsp@ce{\number-\pslly} translate
}%
}}
\def\putsp@ce#1{#1 }
\def\psfordvitps{%       From the UNIX TeXPS package, vers.>3.12
%---------------
% Convert a dimension into the number \psn@sp (in scaled points)
\def\psdimt@n@sp##1{\d@mx=##1\relax\edef\psn@sp{\number\d@mx}}
\def\PSspeci@l##1##2{%
% psfig.psr contains the def of "startTexFig": if you can locate it
% and include the correct pathname, it should work
\special{dvitps: Include0 "psfig.psr"}% contains def of "startTexFig"
\psdimt@n@sp{\drawingwd}
\special{dvitps: Literal "\psn@sp\space"}
\psdimt@n@sp{\drawinght}
\special{dvitps: Literal "\psn@sp\space"}
\psdimt@n@sp{\psllx bp}
\special{dvitps: Literal "\psn@sp\space"}
\psdimt@n@sp{\pslly bp}
\special{dvitps: Literal "\psn@sp\space"}
\psdimt@n@sp{\psurx bp}
\special{dvitps: Literal "\psn@sp\space"}
\psdimt@n@sp{\psury bp}
\special{dvitps: Literal "\psn@sp\space startTexFig\space"}
\special{dvitps: Include1 "##1"}
\special{dvitps: Literal "endTexFig\space"}
}}
\def\psonlyboxes{%     Draft-like behaviour if none of the others works
%---------------
\def\PSspeci@l##1##2{%
\at(0cm;0cm){\boxit{\vbox to\drawinght
  {\vss
  \hbox to\drawingwd{\at(0cm;0cm){\hbox{(##1)}}\hss}
  }}}
}%
}
\def\psloc@lerr#1{%
\let\savedPSspeci@l=\PSspeci@l%
\def\PSspeci@l##1##2{%
\at(0cm;0cm){\boxit{\vbox to\drawinght
  {\vss
  \hbox to\drawingwd{\at(0cm;0cm){\hbox{(##1) #1}}\hss}
  }}}
\let\PSspeci@l=\savedPSspeci@l% restore normal output for other figs!
}%
}
%
%\def\psfor...  add your own!
%
%  \ReadPSize{PSfilename} reads the dimensions of a PostScript drawing
%	     and stores it in \drawinght(wd)
\newread\psiz@
\newdimen\drawinght\newdimen\drawingwd
\newdimen\psxoffset\newdimen\psyoffset
\newbox\drawingBox
\newif\ifNotB@undingBox
\newhelp\PShelp{Proceed: you'll have a 5cm square blank box instead of
your graphics (Jean Orloff).}
\def\@mpty{}
\def\s@tsize#1 #2 #3 #4\@ndsize{
  \def\psllx{#1}\def\pslly{#2}%
  \def\psurx{#3}\def\psury{#4}%  needed by a crazyness of dvips!
  \ifx\psurx\@mpty\NotB@undingBoxtrue% this is not a valid one!
  \else
    \drawinght=#4bp\advance\drawinght by-#2bp
    \drawingwd=#3bp\advance\drawingwd by-#1bp
%  !Units related by crazy factors as bp/pt=72.27/72 should be BANNED!
  \fi
  }
\def\sc@nline#1:#2\@ndline{\edef\p@rameter{#1}\edef\v@lue{#2}}
\def\g@bblefirstblank#1#2:{\ifx#1 \else#1\fi#2}
\def\psm@keother#1{\catcode`#112\relax}% borrowed from latex
\def\execute#1{#1}% Seems stupid, but cs are identified BEFORE execution
{\catcode`\%=12
\xdef\B@undingBox{%%BoundingBox}
}  		%% is not a true comment in PostScript, even if % is!
\def\ReadPSize#1{
 \edef\PSfilename{#1}
 \openin\psiz@=#1\relax
 \ifeof\psiz@ \errhelp=\PShelp
   \errmessage{I haven't found your postscript file (\PSfilename)}
   \psloc@lerr{was not found}
   \s@tsize 0 0 142 142\@ndsize
   \closein\psiz@
 \else
   \loop
     \execute{\begingroup
       \let\do\psm@keother
       \dospecials
       \catcode`\ =10
       \catcode`\^^M=9
       \global\read\psiz@ to\n@xtline
       \endgroup}
     \ifeof\psiz@
       \errhelp=\PShelp
       \errmessage{(\PSfilename) is not an Encapsulated PostScript File:
           I could not find any \B@undingBox: line.}
       \edef\v@lue{0 0 142 142:}
       \psloc@lerr{is not an EPSFile}
       \NotB@undingBoxfalse
     \else
       \expandafter\sc@nline\n@xtline:\@ndline
       \ifx\p@rameter\B@undingBox\NotB@undingBoxfalse
         \edef\int@rmediateresult{%
           \expandafter\g@bblefirstblank\v@lue\space\space\space}
         \expandafter\s@tsize\int@rmediateresult\@ndsize
       \else\NotB@undingBoxtrue
       \fi
     \fi
   \ifNotB@undingBox\repeat
   \closein\psiz@
 \fi
\message{#1}
}
%
% \psboxto(xdim;ydim){psfilename}: you specify the dimensions and
%    TeX uniformly scales to fit the largest one. If xdim=0pt, the
%    scale is fully determined by ydim and vice versa.
%    Notice: psboxes are a real vboxes; couldn't take hbox otherwise all
%    indentation and all cr's would be interpreted as spaces (hugh!).
%
\newcount\xscale \newcount\yscale \newdimen\pscm\pscm=1cm
\newdimen\d@mx \newdimen\d@my
\let\ps@nnotation=\relax
\def\psboxto(#1;#2)#3{\vbox{
   \ReadPSize{#3}
   \divide\drawingwd by 1000
   \divide\drawinght by 1000
   \d@mx=#1
   \ifdim\d@mx=0pt\xscale=1000
         \else \xscale=\d@mx \divide \xscale by \drawingwd\fi
   \d@my=#2
   \ifdim\d@my=0pt\yscale=1000
         \else \yscale=\d@my \divide \yscale by \drawinght\fi
   \ifnum\yscale=1000
         \else\ifnum\xscale=1000\xscale=\yscale
                    \else\ifnum\yscale<\xscale\xscale=\yscale\fi
              \fi
   \fi
   \divide \psxoffset by 1000\multiply\psxoffset by \xscale
   \divide \psyoffset by 1000\multiply\psyoffset by \xscale
   \global\divide\pscm by 1000
   \global\multiply\pscm by\xscale
   \multiply\drawingwd by\xscale \multiply\drawinght by\xscale
   \ifdim\d@mx=0pt\d@mx=\drawingwd\fi
   \ifdim\d@my=0pt\d@my=\drawinght\fi
   \message{scaled \the\xscale}
 \hbox to\d@mx{\hss\vbox to\d@my{\vss
   \global\setbox\drawingBox=\hbox to 0pt{\kern\psxoffset\vbox to 0pt{
      \kern-\psyoffset
      \PSspeci@l{\PSfilename}{\the\xscale}
      \vss}\hss\ps@nnotation}
   \global\ht\drawingBox=\the\drawinght
   \global\wd\drawingBox=\the\drawingwd
   \baselineskip=0pt
   \copy\drawingBox
 \vss}\hss}
  \global\psxoffset=0pt
  \global\psyoffset=0pt% These are local to one figure
  \global\pscm=1cm
  \global\drawingwd=\drawingwd
  \global\drawinght=\drawinght
}}
%
% \psboxscaled{scalefactor*1000}{PSfilename} allows to bypass the
%   rounding errors of TeX integer divisions for situations where the
%   TeX box should fit the original BoundingBox with a precision better
%   than 1/1000.
%
\def\psboxscaled#1#2{\vbox{
  \ReadPSize{#2}
  \xscale=#1
  \message{scaled \the\xscale}
  \divide\drawingwd by 1000\multiply\drawingwd by\xscale
  \divide\drawinght by 1000\multiply\drawinght by\xscale
  \divide \psxoffset by 1000\multiply\psxoffset by \xscale
  \divide \psyoffset by 1000\multiply\psyoffset by \xscale
  \global\divide\pscm by 1000
  \global\multiply\pscm by\xscale
  \global\setbox\drawingBox=\hbox to 0pt{\kern\psxoffset\vbox to 0pt{
     \kern-\psyoffset
     \PSspeci@l{\PSfilename}{\the\xscale}
     \vss}\hss\ps@nnotation}
  \global\ht\drawingBox=\the\drawinght
  \global\wd\drawingBox=\the\drawingwd
  \baselineskip=0pt
  \copy\drawingBox
  \global\psxoffset=0pt
  \global\psyoffset=0pt% These are local to one figure
  \global\pscm=1cm
  \global\drawingwd=\drawingwd
  \global\drawinght=\drawinght
}}
%
%  \psbox{PSfilename} makes a TeX box having the minimal size to
%      enclose the picture
\def\psbox#1{\psboxscaled{1000}{#1}}
%
% \centinsert{anybox} is just a centered \midinsert, but is included as
%    people barely use the original inserts from TeX.
%
\def\centinsert#1{\midinsert\line{\hss#1\hss}\endinsert}
\def\psannotate#1#2{\def\ps@nnotation{#2\global\let\ps@nnotation=\relax}#1}
\def\pscaption#1#2{\vbox{
   \setbox\drawingBox=#1
   \copy\drawingBox
   \vskip\baselineskip
   \vbox{\hsize=\wd\drawingBox\setbox0=\hbox{#2}
     \ifdim\wd0>\hsize
       \noindent\unhbox0\tolerance=5000
    \else\centerline{\box0}
    \fi
}}}
% for compatibility with older versions
\def\psfig#1#2#3{\pscaption{\psannotate{#1}{#2}}{#3}}
\def\psfigurebox#1#2#3{\pscaption{\psannotate{\psbox{#1}}{#2}}{#3}}
%
% \at(#1;#2)#3 puts #3 at #1-higher and #2-right of the current
%    position without moving it (to be used in annotations).
\def\at(#1;#2)#3{\setbox0=\hbox{#3}\ht0=0pt\dp0=0pt
  \rlap{\kern#1\vbox to0pt{\kern-#2\box0\vss}}}
%
% \gridfill(ht;wd) makes a 1cm*1cm grid of ht by wd whose lower-left
%   corner is the current point
\newdimen\gridht \newdimen\gridwd
\def\gridfill(#1;#2){
  \setbox0=\hbox to 1\pscm
  {\vrule height1\pscm width.4pt\leaders\hrule\hfill}
  \gridht=#1
  \divide\gridht by \ht0
  \multiply\gridht by \ht0
  \gridwd=#2
  \divide\gridwd by \wd0
  \multiply\gridwd by \wd0
  \advance \gridwd by \wd0
  \vbox to \gridht{\leaders\hbox to\gridwd{\leaders\box0\hfill}\vfill}}
%
% Useful to measure where to put annotations
\def\fillinggrid{\at(0cm;0cm){\vbox{
  \gridfill(\ht\drawingBox;\wd\drawingBox)}}}
%
% \textleftof\anybox: Sample text\endtext
%   inserts "Sample text" on the left of \anybox ie \vbox, \psbox.
%   \textrightof is the symmetric (not documented, too uggly)
% Welcome any suggestion about clean wraparound macros from
%   TeXhackers reading thisTeXhackers reading this
%
\def\textleftof#1:{
  \setbox1=#1
  \setbox0=\vbox\bgroup
    \advance\hsize by -\wd1 \advance\hsize by -2em}
\def\textrightof#1:{
  \setbox0=#1
  \setbox1=\vbox\bgroup
    \advance\hsize by -\wd0 \advance\hsize by -2em}
\def\endtext{
  \egroup
  \hbox to \hsize{\valign{\vfil##\vfil\cr%
\box0\cr%
\noalign{\hss}\box1\cr}}}
%
% \frameit{\thick}{\skip}{\anybox}
%    draws with thickness \thick a box around \anybox, leaving \skip of
%    blank around it. eg \frameit{0.5pt}{1pt}{\hbox{hello}}
% \boxit{\anybox} is a shortcut.
\def\frameit#1#2#3{\hbox{\vrule width#1\vbox{
  \hrule height#1\vskip#2\hbox{\hskip#2\vbox{#3}\hskip#2}%
        \vskip#2\hrule height#1}\vrule width#1}}
\def\boxit#1{\frameit{0.4pt}{0pt}{#1}}
\catcode`\@=12 % cs containing @ are unreachable
%
% CUSTOMIZE YOUR DEFAULT DRIVER:
%    Uncomment the line corresponding to your TeX system:
%\psfortextures%     For TeXtures on the Macintosh
%\psforoztex   %     For OzTeX shareware on the Macintosh
%\psfordvitops %     For the DVItoPS converter for TeX on IBM mainframes
 \psfordvips   %     For DVIPS converter on VAX and UNIX
%\psfordvitps  %     For dvitps from TeXPS package under UNIX
%\psonlyboxes  %     Blank Boxes (when all else fails).
%----------------------------end ofpsbox.tex & beginning of ewb.tex

\begin{figure}
$$ \psboxscaled{150}{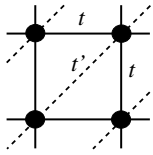} $$
\caption{Lattice structure of tt'-TH model on triangular lattice with anisotropic transfers
$t$ and $t'$.}
\label{Fig1}
\end{figure}

In particular, the case between $t'/t \sim 0.5$ and $1.0$ has attracted a particular interest
because tight binding fit of the band structure by the extended H\"uckel approximation
for $\kappa$-type BEDT-TTF
compound suggests that this region may provide a minimal model after eliminating 
the internal degrees of freedom in the dimerized two ET molecules~\cite{Mori}.  
In terms of clarifying the metal-insulator (MI) transition and the accompanied superconductivity,
the ground state of this 
model has been a subject of intensive studies but mainly on the Hartree-Fock or 
FLEX levels~\cite{Mackenzie,Kino,Moriya,Kuroki}.
Because significant correlation and quantum effects
are expected as described above, it is desired to 
clarify the phase diagram of the basic $t$-$t'$-TH models itself by fully taking 
account of quantum fluctuations.

Our result shows remarkably a generic emergence of nonmagnetic insulating state
sandwitched by a Mott MI transition and an AF transition 
in contrast with the previous mean-field predictions, while
a nonmagnetic insulating state is also suggested in different Hubbard models on the square lattice
with next-nearest neighbor transfers~\cite{KashimaImadaMIT}.
We also find that the nonmagnetic insulators do not show simple symmetry 
breakings such as dimer, plaquette, and flux phases, in support for the appearance of
genuine Mott insulator distinct from band insulators. 

The genuine Mott insulating state not adiabatically connected to the band insulator
has been a subject of long-standing search since Anderson's proposal~\cite{Anderson}. 
This proposed RVB state is in basic sense realized in the 1D systems
%with Tomonaga-Luttinger type criticality of spin degrees of freedom
, while it remains a challenge 
for multi-dimensional systems.  We stress that such 
RVB state may be found more easily near the bandwidth-control metal to Mott insulator 
transition, where quantum mechanical fluctuations of spin as well as charge destroy trivial 
symmetry breakings and the local moment is relatively small.  
In the light of our results, we argue that this category of genuine Mott insulator 
may indeed be realized in $\kappa$-(ET)$_2$Cu$_2$(CN)$_3$, which seems to be nonmagnetic and insulating at 
ambient pressure down to 1.4K.

We introduce $t$-$t'$-TH model on a triangular lattice shown in Fig.~\ref{Fig1} with the 
Hubbard Hamiltonian 
$H= -\sum_{\langle i,j \rangle ,\sigma}t   
\left(c_{i\sigma}^{\dagger}c_{j\sigma}+{\rm H.c.}\right)
+ \sum_{\langle k,l \rangle,\sigma}t' 
\left(c_{k\sigma}^{\dagger}c_{l\sigma}+{\rm H.c.}\right)
+ U\sum_{i=1}^{N}   n_{i\uparrow}n_{i\downarrow}$
%$H=H_t+H_U$ with $H_t= -\sum_{\langle i,j \rangle ,\sigma}t   
%\left(c_{i\sigma}^{\dagger}c_{j\sigma}+{\rm H.c.}\right)
%+ \sum_{\langle k,l \rangle,\sigma}t' 
%\left(c_{k\sigma}^{\dagger}c_{l\sigma}+{\rm H.c.}\right)$
%and $H_{U}=U\sum_{i=1}^{N}   n_{i\uparrow}n_{i\downarrow}$
%\begin{eqnarray}  H&=&\hat{H}_{t}+\hat{H}_{U},\nonumber\\  
%\hat{H}_{t}=-\sum_{\langle i,j \rangle ,\sigma}t   
%\left(c_{i\sigma}^{\dagger}c_{j\sigma}+H.c.\right),\nonumber\\ 
%&&\    + \sum_{\langle k,l \rangle,\sigma}t' 
%\left(c_{k\sigma}^{\dagger}c_{l\sigma}+H.c.\right),\nonumber\\ 
%\hat{H}_{U}&=&U\sum_{i=1}^{N}   \left(n_{i\uparrow}-\frac{1}{2}\right)   
%\left(n_{i\downarrow}-\frac{1}{2}\right) ,
%\label{Hamiltonian}  
%\end{eqnarray} 
with the standard notation.  
%We take $t=1.0$ as the energy scale. 
In this letter, we apply the recently developed 
PIRG method~\cite{KashimaImadaPIRG} 
to clarify the ground state in the plane of $U/t$ and $t'/t$.  
This algorithm allows us to start from and improve a mean-field Hartree-Fock solution and 
reach the ground-state by taking account 
quantum fluctuations in a systematic fashion.
By increasing the dimension of truncated Hilbert space in a nonorthogonal 
basis numerically optimized by the path-integral operation, a variance extrapolation is taken to 
reach the true ground state of finite size systems in a 
controlled way. We took number of Slater basis functions up to around 300. This numerical method does not have difficulties such as the negative sign problem in the quantum Monte Carlo calculations.
In the present study, we take system size up to $N=196$ sites ($14 \times 14$) with the periodic boundary
condition.  Using the results at various sizes, we extrapolate to the 
thermodynamic limit using appropriate scaling procedure and eliminate finite size effects.

To see the MI transition, we calculate the singularity of the ground-state 
energy $E_g$ as a function of the control parameter, $U/t$. The level crossing with a 
first-order transition or singularity with a continuous character 
is detected as a signature of the MI transition
with complementary analysis of the charge gap and the momentum distribution.
The AF transition is studied by equal-time spin correlation 
functions in the momentum space.  We also calculate several density wave correlations
to discuss possible symmetry breakings such as staggered flux as well as dimer correlations. 

\begin{figure}
$$ \psboxscaled{600}{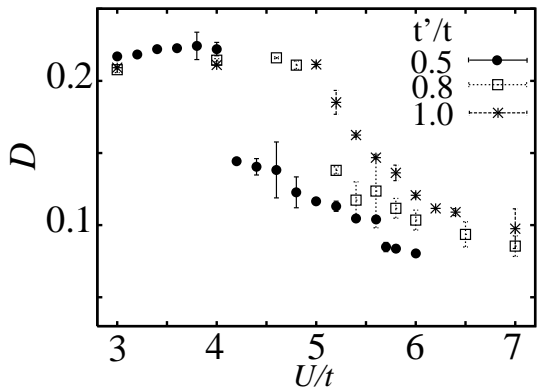} $$
\caption{Averaged double occupancy $D$ as a function of $U/t$ for $t'/t=0.5$(filled circle),
0.8(open square) and 1.0(asterisk). The plotted error bars are only from the size extrapolation.
The errors produced by the variance extrapolation[2] in 
PIRG is not shown and are at most 0.02.}
\label{double_occupancy}
\end{figure}

In Fig.~\ref{double_occupancy}, we show $\partial E_{g}/\partial U$ as a function of $U$.
Jumps of the averaged double occupancy $D = \langle n_{i\uparrow 
}n_{i\downarrow} \rangle = \partial E_{g}/\partial U /N+1/4$ signal first-order MI transitions 
induced by level crossings at $U/t=4.1 \pm 0.1$, and $5.0 \pm 0.2$ for 
choices $t'/t=0.5$, and $0.8$, respectively, while a continuous transition is plausible 
at $5.2 \pm 0.2 $ for $t'/t=1.0$.  
These transitions show similar character to MI transitions identified by charge gap formation 
and qualitative change of the momentum distributions for other 
types of Hubbard models~\cite{KashimaImadaMIT}.  Fig.~\ref{momdist}
exemplifies that the momentum distributions of 
the $t$-$t'$-TH model indeed show transition from a typical 
metallic behavior to that of an insulator across this boundary .

\begin{figure}
$$ \psboxscaled{400}{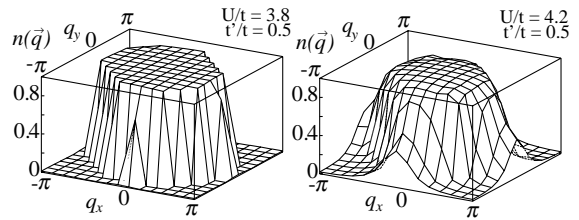} $$
\caption{Momentum distribution on 14 $\times$ 14 lattice at $t'/t=0.5$ for $U/t=3.8$ and 4.2.}
\label{momdist}
\end{figure}

The magnetic transition is probed by the Fourier transform of the equal-time spin correlation, 
$S(\mib{q})$ defined by 
$ S\left(\mib{q}\right)=\frac{1}{3N}\sum_{i,j}^{N} 
  \left\langle\mib{S}_{i}\cdot \mib{S}_{j}
  \right\rangle e^{i\mib{q}\cdot \left(\mib{R}_{i}-\mib{R}_{j}\right)}$,
%\begin{eqnarray}
% S\left(\mib{q}\right)=\frac{1}{3N}\sum_{i,j}^{N} 
%  \left\langle\mib{S}_{i}\cdot \mib{S}_{j}
%  \right\rangle e^{i\mib{q}\cdot \left(\mib{R}_{i}-\mib{R}_{j}\right)},
%  \label{spinco}
%\end{eqnarray}
where $\mib{S}_{i}$ is the spin of the $i$-th site with $\mib{R}_{i}$ 
representing the coordinate of the $i$-th site.
At $t'/t=0.5$ the spin correlation shows sharp commensurate
paek at $(\pi,\pi)$ for $U/t \geq 5.7$ while it remains rather small and incommensurate 
for $U/t \leq 5.6$. The size scaling in Fig.~\ref{spinfss} leads to $U$ dependence of the 
staggered moment $m=\sqrt{\lim_{N \rightarrow \infty}S(\mib{q_{peak}})/N}$ as in Fig.~\ref{stagmag},
where a first-order (or very sharp continuous) AF transition is identified 
at $U/t=5.65 \pm 0.05$. A small jump of $\partial E_{g}/\partial U$ seen in Fig.~\ref{double_occupancy} at this 
point also supports the first-order character.  For larger values of $t'/t$, at $t'/t=0.8$
and 1.0, the size scaling of the spin correlation shows the absence of magnetic symmetry breaking
at all momenta and all $U$ studied, namely for $U/t \le 10$.   

\begin{figure}
$$ \psboxscaled{550}{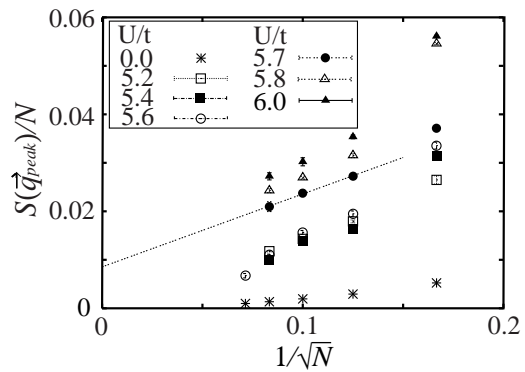} $$
\caption{Size scaling for the peak value of the equal-time spin structure factor $S(q_{peak})$ at $t'/t=0.5$.}
\label{spinfss}
\end{figure}

\begin{figure}
$$ \psboxscaled{550}{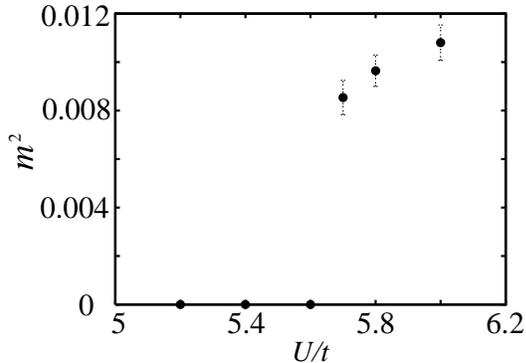} $$
\caption{Staggered magnetization in the thermodynamic limit as a function of $U/t$ at $t'/t=0.5$.}
\label{stagmag}
\end{figure}

The phase diagram of the $t$-$t'$-TH model constructed from our calculations is illustrated in 
Fig.~\ref{phase_diag}.  A remarkable feature is the emergence of nonmagnetic insulating phase sandwitched by
MI and AF transitions. This is in contrast with the Hartree-Fock 
results where the MI transition occurs at larger $U/t$ than the AF transition.  The
 triangular lattices were also studied in the spin-1/2 Heisenberg model corresponding to the limit 
$U\rightarrow \infty$~\cite{triangular1,triangular2}.  
Although the decisive conclusion cannot be drawn due to the lack 
of powerful techniques, it was claimed from a small cluster studies
or series expansions that either long-range AF order or dimer order appear in this model on a regular triangular lattice.  This suggests a possibility that the reduction of
the local moment due to charge fluctuations at $U$ reduced to a finite value drives the destruction of the order.
In fact, for example, the local moment $m_l=\sqrt{\langle n \rangle -2D}/2$ is reduced to 
$m_l\sim 0.4$ at the phase boundary $U/t=5.2$ at $t'/t=1.0$.
\begin{figure}
$$ \psboxscaled{550}{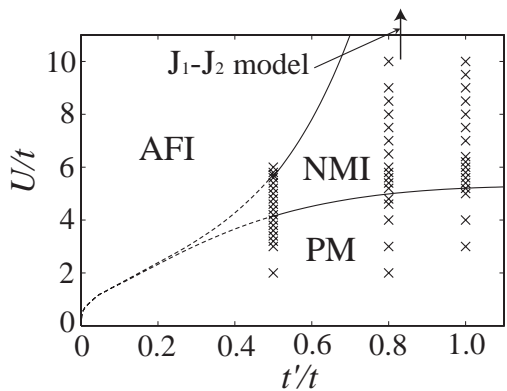} $$
%$$ \psboxscaled{450}{phasediagz.eps} $$
\caption{Phase diagram of $t$-$t'$ TH model in the plane of $U/t$ at $t'/t$. AFI,NMI and PM 
represent AF insulator, nonmagnetic insulator, and paramagnetic metal, respectively. 
Calculations have been done at the cross points.  The arrow shows speculated boundary from AFI to the dimer insulator in the 
limit $U \rightarrow \infty$ for the $J_1$-$J_2$ model~\cite{triangular2}.}
\label{phase_diag}
\end{figure}
   
Insulators with some translational symmetry breakings such as the AF and 
dimer orders may be adiabatically continued to band insulators due to the band gap
formation through extension of the unit cell and the folded Brillouin zone, and hence  do not
belong to the genuine Mott insulator discussed above.  We first have to note that it is hard to prove the existence 
of the genuine Mott
insulator because it is 
difficult to prove the absence of all the possible translational symmetry breakings.
Under this circumstance, an available way to get insight on this problem is to examine
every symmetry breaking ever proposed and judge the plausibility of such a phase. 

Along this line of motivation, we have further calculated 
correlations for several symmetry breakings claimed in the literature 
such as dimer order~\cite{triangular2,Sachdev}, plaquette 
order~\cite{Sorellaplaquette} and various density waves~\cite{Affleck} in this remarkable nonmagnetic insulating phase.  
Here the density wave correlation is defined by
 $ D_{\alpha}(\mib{q})=
  \left\langle\Delta_{\alpha}^{\dagger}(\mib{q})\Delta_{\alpha}(\mib{q})
  \right\rangle $
% \begin{eqnarray}
% D_{\alpha}(\mib{q})=
%  \left\langle\Delta_{\alpha}^{\dagger}(\mib{q})\Delta_{\alpha}(\mib{q})
%  \right\rangle 
%  \label{DWCR}
%\end{eqnarray}
where 
$\Delta_{\alpha}^{\dagger}(\mib{q})=\sum_{\mib{k},\sigma}f_{\alpha}(\mib{k})
c^{\dagger}_{\mib{k} \sigma}c_{\mib{k}+\mib{q},\sigma}$.
% \begin{eqnarray}
% \Delta_{\alpha}^{\dagger}(i)=\sum_{\mib{k}}f_{\alpha}(\mib{k})
%c^{\dagger}_{\mib{k} \uparrow}c_{\mib{k} \downarrow}
%e^{i\mib{k}\cdot\mib{R}_{i}}
%  \label{DWOP}
%\end{eqnarray}
We have studied the cases $f_{1s}(\mib{k})=\cos k_x + \cos k_y$, 
$f_{2s}(\mib{k})=2\cos k_x \cos k_y$, 
$f_{1d}(\mib{k})=\cos k_x - \cos k_y$, and $f_{2d}(\mib{k})=2 \sin k_x \sin k_y$.
We note the case $1d$ represents the order parameter for the staggered flux 
(or $d$-density wave) state~\cite{Affleck}. 
Fig.~\ref{DDW} shows a typical size scalings of these density wave correlations for the $t$-$t'$-TH model. Other types show similar behaviors.
It does not seem to show symmetry breakings and the correlation actually
remains rather short ranged in all the parameter space we have studied ($0 \leq U/t \leq 10$
and $0.5 \leq t'/t \leq 1.0$).
The possibility of dimer, plaquette and density wave orders have also been examined in the 
Hubbard model on a square lattice with diagonally crossing transfers between next-nearest 
neighbor sites.
The size scalings again suggest the absence of these translational symmetry breakings~\cite{Watanabe}.
Although these results do not exclude a possiblity for the translational symmetry breakings other 
than those studied here, we conclude that the theoretically proposed phases so far are unlikely
to be realized in our phase hence the existence of genuine Mott insulating state becomes
much more plausible.

\begin{figure}
$$ \psboxscaled{500}{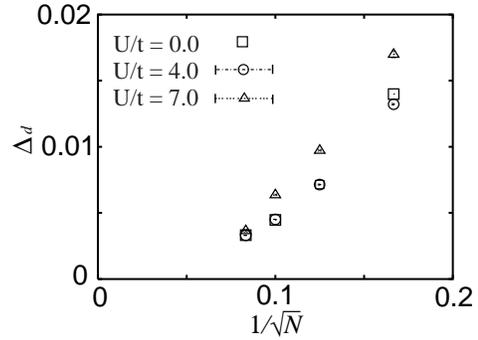} $$
\caption{Size scaling for the d density wave (stuggered flux) correlation $\Delta_d=
\vert D_{1d}({\mib q}_{peak})\vert/N$ for $t'/t=1.0$.}
\label{DDW}
\end{figure}

Our result leads us 
to an intriguing view for recent experimental results on organic compounds. 
As we discussed above, the $t$-$t'$-TH model provides a simplest effective Hamiltonian for
the $\kappa$-type ET compounds if we may take the dimerization of two parallel ET molecules
large~\cite{Mori}.  The HOMO band is split by the dimerization
and the upper band provides the effective Hubbard model at half filling.  
The effective interaction for $U$ in this case is taken as 
$U_{eff}=2t_{b1} + \frac{U_{ET}}{2}[1-\sqrt{1+(\frac{4t_{b1}}{U_{ET}})^2}]$,
where $U_{ET}$ is the original onsite Coulomb repulsion for the ET molecule
and $t_{b1}$ is the intradimer transfer.  

Among available compounds, the band structure calculation~\cite{Kobayashi,SaitoG} 
suggests $\kappa$-(ET)$_2$Cu$_2$(CN)$_3$ has the largest ratio of $t'/t \sim 1.0$ 
($t \sim 0.055$ eV and $t' \sim 0.058$ eV ) with 
rather large $U_{eff}/t \sim 8.2$.  
This compound is close to the metal-insulator phase boundary because
a small pressure around 0.05GPa drives the insulating to the
superconducting phase.  
The ESR measurement of $\kappa$-(ET)$_2$Cu$_2$(CN)$_3$ at ambient pressure as well as 
the susceptibility
seems to suggest that the insulating phase remains nonmagnetic with broad 
structure of the spin susceptibility~\cite{SaitoG}.  Very recent NMR data~\cite{Kanoda}
 also suggest
that the $1/T_1$ remains constant above 100K while it starts substantially decreasing below
50K and there exists no signature of the magnetic transition down to 1.4K so far.  
No indication of lattice distortion has been reported.

It is very attractive to relate these experimental results to our calculated phase diagram
with the insulating phase without symmetry breakings, because the parameter values 
deduced from the band structure calculation correspond to $t'/t=1.0$ where
we obtain large nonmagnetic insulating phase.  
In other compounds such as deuterated $\kappa$-(ET)$_2$Cu[N(CN)$_2$]Br, 
the first-order transition between a superconductor and an AF
insulator is observed.  These compounds have 
relatively small parameters $t'/t \sim 0.5 - 0.6$~\cite{Fortunelli}.  In our result at $t'/t=0.5$, there
still exists nonmagnetic insulating phase.  This discrepancy is speculated to be due to
an oversimplification to the effective single-band model.  In fact, internal degrees of freedom
of dimer stabilizes metals more than insulator, while the geometrical frustration
is relaxed by an assymetric distribution of electrons on a dimer and the AF order  
may not be suppressed strongly, which results in shrink of the nonmagnetic insulator phase.
Necessity of multi-band models was also pointed out in different contexts~\cite{Schmalian,KurokiMatsuda}.
Quantitative analysis of this multi-band effect is left for further studies. 
If the AF and MI transitions become close, lattice distortion accompanied by the MI transition 
necessarily enhance the first-order nature
%.  This effect
%beyond our scope may also lead to the shrink of the nonmagnetic insulator.
also leading to the shrink of the nonmagnetic insulator.

The absence of strong AF correlations near the MI transition 
implies that the mechanism of the superconductivity seen in 
$\kappa$-(ET)$_2$Cu$_2$(CN)$_3$ under pressure may not be directly connected with AF 
fluctuations.
In our PIRG calculation, although the pairing correlations definitely remain short-ranged
for $U/t=3$ at $t'/t=0.5$ and for $U/t=4$ at $t'/t=0.8$ and 1.0, 
the d-wave pairing correlations near the MI boundary does not show a good 
convergence 
which indicates the necessity of larger number of Slater determinants in PIRG for this quantity to get 
converged results, implying large quantum fluctuations. This will be discussed elsewhere.

We also point out, however, a fundamentally interesting region still waits for experimental realization:
In ET dimerized compounds, stronger dimerization is required. 
%If effective $t'/t$ and $U/t$ can be taken large for systems modeled by this 
%$t$-$t'$-TH Hamiltonian, we expect stabilization of the genuine Mott insulator 
%without simple symmetry breakings and the MI transition becomes rather continuous.  
%This may correspond to stronger dimerization with $t'/t\sim 1$
%and $U/t \sim 5$ in $\kappa$-type BEDT-TTF compounds although 
%it is not clear whether it is realizable.  
More sophisticated realization of nonmagnetic insulator and metals near it in 
the frustrated lattice would be desired also in other materials.  

In summary, we have studied the Hubbard model on anisotropic triangular lattice ($t$-$t'$-TH model).  
The ground state is obtained by the PIRG method,
where the Hartree-Fock results are systematically improved to reach the true ground state.
The phase boundary of bandwidth control MI and AF
transitions in the parameter space of $t'/t$ and $U/t$ shows that
a nonmagnetic insulator appears between these two transitions.
Simple translational symmetry breakings such as flux phases remain short ranged and genuine
Mott insulating state seems to be realized.  We have pointed out that $\kappa$-(ET)$_2$Cu$_2$(CN)$_3$ 
may belong to this category of the genuine Mott insulator within the available data.  
The degeneracy of the ground state as well as the spin excitation gap in the nonmagnetic insulator phase is an intriguing 
future problem to be clarified. 
   
%\section*{Acknowledgments} 
We thank K. Kanoda for valuable discussions.  
The work is supported by the `Research for the Future' program from JSPS under grant number JSPS-RFTF97P01103. A part of the computation was done at the supercomputer center in ISSP, University of Tokyo.   
%\input{bib.tex}  }
%%%%%% 

\end{document}